\documentclass[prd,aps]{revtex4}
\usepackage{amsmath}
\usepackage{amssymb}
\usepackage{amsfonts}
\usepackage{graphicx,bm}
\usepackage{dcolumn}
\usepackage[colorlinks=true]{hyperref}
\usepackage{epsf}
\usepackage{enumerate}
\usepackage{hhline}
\usepackage{array}
\usepackage{tabularx}
\usepackage{subfigure}

\newcommand{\be}{\begin{equation}}
\newcommand{\ee}{\end{equation}}
\newcommand{\bea}{\begin{eqnarray}}
\newcommand{\eea}{\end{eqnarray}}
\newcommand{\beaa}{\begin{eqnarray*}}
\newcommand{\eeaa}{\end{eqnarray*}}

\newcommand{\nn}{\nonumber \\}
\newcommand{\e}{\mathrm{e}}

\def\be{\begin{equation}}
\def\ee{\end{equation}}
\def\bea{\begin{eqnarray}}
\def\eea{\end{eqnarray}}

\begin{document}
\title{Constant roll inflation in multifield models}

\author{Merce Guerrero} \email{merguerr@ucm.es}
\affiliation{Departamento de F\'isica Te\'orica and IPARCOS,
	Universidad Complutense de Madrid, E-28040 Madrid, Spain}

\author{Diego Rubiera-Garcia} \email{drubiera@ucm.es}
\affiliation{Departamento de F\'isica Te\'orica and IPARCOS,
	Universidad Complutense de Madrid, E-28040 Madrid, Spain}

\author{Diego S\'aez-Chill\'on G\'omez}\email{diego.saez@uva.es}
\affiliation{Departmento de F\'isica Te\'orica, At\'omica y \'Optica, Campus Miguel Delibes, \\ University of Valladolid UVA, Paseo Bel\'en, 7,
47011 - Valladolid, Spain}

\begin{abstract}

Constant roll inflation is analyzed in the presence of multi scalar fields which are assumed to be described by a constant roll rate each. The different cases are studied and the corresponding potentials are reconstructed. The exact solutions are obtained, which show a similar behaviour to the single scalar field model. For one of the cases analyzed in the paper, the so-called adiabatic field also constant rolls while entropy perturbations become null, while the second case may lead to non-adiabatic perturbations. Both cases can fit well the Planck data by assuming the appropriate values for the free parameters of the models.

\end{abstract}
%
%
\maketitle
%
%
%
\section{Introduction}

Since cosmic inflation was proposed in order to sort out some of the problems inherent to the Big Bang model, a lot of literature has been published about this paradigm, with many theoretical models that are capable of reproducing a super-accelerating phase just after the Big Bang singularity that solve the initial conditions problem in the standard cosmological model \cite{reviews}. In addition, the release of data by the missions WMAP first, and Planck later, made possible to test some features of inflation through the imprints on the anisotropies in the Cosmic Microwave Background (CMB) \cite{WMAP,Akrami:2018odb,Ade:2015lrj}. This is due to the fact that inflation not only can solve the initial conditions problem of the Big Bang model, but also the quantum fluctuations produced during this early period yield the anisotropies in the CMB which are the seeds causing the variations on the matter distribution that later formed the galaxies and clusters of galaxies \cite{Mukhanov:1990me}.

Most of the inflationary models are constructed by means of a single scalar field, the so-called inflaton, whose potential is chosen in such a way that inflation is produced on a plateau of the potential leading to a quasi de Sitter expansion, and then the field goes down up to a minimum, where inflation ends. Such models typically  assume that the scalar field slow rolls down the slope of the potential, during which fluctuations of the scalar field produce scalar and tensor perturbations, that can be related to the potential of the inflaton. This yields a value for the spectral index for curvature fluctuations and the ratio between tensor and scalar perturbations, magnitudes that can be compared to the data from CMB, leading to constraints on the shape of the scalar potential \cite{Liddle:1994dx,Lidsey:1995np}. Nevertheless, inflation can be also well produced with other frameworks different than a scalar field leading to the same accurate predictions. Particularly, inflation have been widely analyzed in the context of some extensions of General Relativity, as in the so-called $f(R)$ gravities \cite{Cognola:2007zu}, with the Starobinsky model as the most promising one for its good predictions at all levels \cite{Starobinsky:1980te}. However, such modified gravity models can be reduced to a single field model, what implies at least a mathematical equivalence among them. Nevertheless, when more than a scalar field is considered, the presence of at least a second field may produce non-adiabatic (isocurvature) perturbations, which  contribute as a source for the adiabatic ones, also at super-horizon scales, and consequently to modifications on the predictions from inflation \cite{GarciaBellido:1995qq,Kaiser:2013sna,Gordon:2000hv,Bassett:2005xm}, although such deviations may be small or null when the fields behave similarly in the field space or when the initial non-adiabatic perturbations are small \cite{Weinberg:2004kf}. 

However, in all the above scenarios the slow roll condition on the scalar field(s) is assumed in general, which basically means to consider a negligible acceleration for the field(s). Nevertheless, inflationary models beyond the slow roll condition have been considered in the literature, as in the ultra slow roll inflation \cite{Kinney:2005vj,Namjoo:2012aa,Martin:2012pe}, where curvature perturbations are not kept frozen at super Hubble scales, inducing non-gaussianities in the power spectrum. All these models have been generalized under the so-called constant roll inflation, which assumes that the rate between the acceleration and velocity of the inflaton remains constant, being the other scenarios just particular cases of this one \cite{Motohashi:2014ppa}. In addition, some exact solutions have been obtained and the corresponding potentials for the scalar field reconstructed, showing that one of these potentials does not induce evolution of the curvature perturbations at superhorizon scales, leading to a viable model for inflation that also satisfies the last constraints from Planck \cite{Motohashi:2017aob,GalvezGhersi:2018haa}. Moreover,  constant roll inflation has been also analyzed in contexts beyond the single scalar field model, as in Brans-Dicke like theories \cite{Motohashi:2019tyj}, modified gravities, as $f(R)$ and $f(T)$ gravity \cite{Nojiri:2017qvx}, or with couplings to gauge fields \cite{Ito:2017bnn}. In addition, some generalizations of the constant roll condition have been studied in \cite{Cicciarella:2017nls} as well as transitions between slow and constant roll scenarios \cite{Odintsov:2017yud}. Also more accurate methods for calculating the power spectrum of scalar and tensor perturbations in constant roll inflation have been proposed \cite{Motohashi:2017gqb,Yi:2017mxs}. 

The main aim of the present paper is to  extend the analysis of constant roll inflation to the presence of more than one scalar field, applying the previous knowledge on multifield inflationary models to the case of two scalar fields that both hold the constant roll condition. In Ref.~\cite{Micu:2019fju}, a two scalar fields model has been studied by considering the constant roll condition on the Hubble parameter and its derivatives and then reconstructing the corresponding solutions for the scalar fields and its potential. Here we assume two separate constant roll conditions, one for each scalar field, and analyze the solutions and possible potentials for the different cases that arise in the model. We shall show that in the case where both fields have similar constant roll rates, the so-called adiabatic field constant rolls too, a result that is obtained also in \cite{Micu:2019fju} by a different approach, and the non-adiabatic perturbations are null. Then, the corresponding potential is reconstructed and the model is confronted with the data from Planck. Also the general scenario where both fields constant roll differently is studied and its predictions confronted to the Planck data, showing that in both cases the predictions of the models can be compatible with such data by assuming suitable values for the free parameters of the models.

The paper is organized as follows: in section \ref{Constant}, we review constant roll inflation with the presence of a single scalar field. Section \ref{multifields} is devoted to show the main general features and tools used in multifield inflation, while in section \ref{multifieldsConstant} we study a two constant roll scalar fields model and reconstruct the solutions and potentials for different scenarios. Also its predictions and comparison to Planck data is carried out. Finally, section \ref{conclusiones} gathers the results and conclusions of the paper.

\section{Constant roll inflation in single field models}
\label{Constant}

Let us start by reviewing the main features of constant roll inflation with a single scalar field. The gravitational action of a scalar field minimally coupled to gravity reads as:
\be
S=\int d^4x\sqrt{-g} \left(\frac{M_{Pl}^2}{2}R-\frac{1}{2}\partial_{\mu}\phi\partial^{\mu}\phi-V(\phi)\right)\ ,
\label{singlefieldaction}
\ee
where $M_{Pl}^2=(8\pi G)^{-1}$ is the Planck mass, $g$ is the determinant of the space-time metric $g_{\mu\nu}$, $R\equiv g_{\mu\nu}R^{\mu\nu}$ is the Ricci scalar, and $V(\phi)$ is the vector potential. The corresponding FLRW equations in a spatially flat universe $ds^2=-dt^2+a(t)^2dx_i^2$, where $a(t)$ is the scale factor, are given by:
\be
3M_{Pl}^2 H^2=\frac{1}{2}\dot{\phi}+V(\phi)\ , \quad -2M_{Pl}^2 \dot{H}=\dot{\phi}^2\ .
\label{FLRWeqssingle}
\ee
Here $H=\frac{\dot{a}}{a}$ is the Hubble parameter and dots refer to derivatives with respect to the cosmic time. In addition, by varying the action with respect to the scalar field $\phi$, its equation in a FLRW metric yields:
\begin{equation}\label{scalarfield}
\ddot{\phi} + 3H \dot{\phi} + \frac{\partial V}{\partial\phi}=0 \ .
\end{equation}
In slow roll inflation, the universe expansion becomes quasi de Sitter as the scalar field behaves as an effective cosmological constant or, in other words:
\be
\ddot{\phi}<<\dot{\phi}\ , \quad \dot{\phi}^2<<V(\phi)\ .
\label{slowrollconds}
\ee
Then, the accelerating expansion should last a large enough number of e-foldings, usually $N=50-65$, after which the scalar field rolls down the potential slope to a minimum. Hence, the corresponding scalar potential has to be a monotonically decreasing function with a plateau at the top (for an analysis and reconstruction of slow roll inflation potentials, see e.g. \cite{Liddle:1994dx,Lidsey:1995np}). Moreover, the above conditions can be more conveniently expressed in terms of the so-called slow roll parameters:
\be
\epsilon=\frac{M_{Pl}^2}{2}\left(\frac{V'}{V}\right)^2\ ,\quad \eta=M_{Pl}^2\frac{V''}{V}\ .
\label{sllowRollParam}
\ee
As inflation occurs, $\epsilon<<1$ and $\eta<<1$, while at the end of inflation $\epsilon\sim 1$. In addition, during inflation  fluctuations on the scalar field grow with the expansion leading to the fluctuations on the metric and, consequently, on the matter density that forms the seeds of the large scale structure of the universe as well as the anisotropies in the CMB. The relation between the curvature perturbations and the slow roll parameters are given by the so-called spectral index that describe the growth of such perturbations, $n_s$, as well as by the tensor to scalar perturbations ratio, $r$, as
\be
n_s-1=-6\epsilon+2\eta\ , \quad r=16\epsilon\ ,
\label{spectralAndr}
\ee
respectively. The data by Planck provides strong constraints on these magnitudes, such that any inflationary model can be tested at least through the type of curvature and tensor perturbations that produces, which has led to rule out some models \citep{Akrami:2018odb,Ade:2015lrj}.

Instead of assuming the slow roll conditions (\ref{slowrollconds}), one may consider a type of potential that leads to a constant rate of roll for the scalar field, which in terms of its derivatives can be expressed as follows:
\begin{equation}
\dfrac{\ddot{\phi}}{\dot{\phi}} = \beta H\ .
\label{constantRollcond}
\end{equation}
This is the constant roll condition that imprints an alternative dynamic to the scalar field \cite{Motohashi:2014ppa}, where the constant
$\beta$ determines the deviation from a flat potential. When $\beta \simeq 0 $,  slow roll inflation is recovered, whereas $\beta = 0 $ corresponds to ``ultra slow roll" inflation \cite{Kinney:2005vj,Namjoo:2012aa,Martin:2012pe}. In order to reconstruct the appropriate scalar potential that holds the constant roll condition (\ref{constantRollcond}), the second FLRW equation in (\ref{FLRWeqssingle}) is expressed as:
\be
\dot{\phi}=-2M_{Pl}^2\frac{\partial H}{\partial\phi}\ .
\label{dotphiH}
\ee
Together with the condition (\ref{constantRollcond}), the following equation for $H$ as a function of the scalar field $\phi$ is obtained:
\be
\frac{\partial^2 H}{\partial\phi^2}+\frac{\beta}{2M_{Pl}^2}H=0\ ,
\label{eqHphi}
\ee
which can be easily solved to obtain the Hubble factor as
\be
H(\phi)=C_1 \e^{\sqrt{\frac{-\beta}{2}}\frac{\phi}{M_{Pl}}}+C_2 \e^{-\sqrt{\frac{-\beta}{2}}\frac{\phi}{M_{Pl}}}\ .
\label{HubblePhi}
\ee
where $C_{1,2}$ are integration constants. Finally, by using the first FLRW equation (\ref{FLRWeqssingle}), the potential is reconstructed as \cite{Motohashi:2014ppa}
\be
V(\phi)=C_1^2M_{Pl}^2(\beta+3)\e^{\sqrt{-2\beta}\frac{\phi}{M_{Pl}}}+C_2^2M_{Pl}^2(\beta+3)\e^{-\sqrt{-2\beta}\frac{\phi}{M_{Pl}}}+2C_1C_2M_{Pl}^2(3-\beta)\ .
\label{PotentialSingle}
\ee
Depending on the value of $\beta$ and $C_{1,2}$, the potential will be characterized by trigonometric or hyperbolic functions, each possibility leading to a different type of inflation. In particular, for $-1<\beta<0$, the potential can be identified with power law inflation which is excluded from observations, since it predicts a too large tensor to scalar ratio. For the general case $\beta<0$ and choosing a hyperbolic cosine in (\ref{HubblePhi}), the potential leads to a type of inflation that behaves as pressureless matter with a cosmological constant, i.e., the $\Lambda$CDM model, which requires additional assumptions for ending inflation. Hence, the only possibility left that provides a viable model for inflation is $\beta>0$, and in such a case the Hubble parameter (\ref{HubblePhi}) turns out:
\be
H(\phi)=M \cos\left(\sqrt{\frac{\beta}{2}}\frac{\phi}{M_{Pl}}\right)\ ,
\label{HubblePhi1}
\ee
being $ M $ a combination of $ C_{1,2} $. The scalar potential becomes in this case:
\be
V(\phi) = 3 M^2 M^2_{pl} \left[1 - \dfrac{3+\beta}{6} \left\lbrace 1 - \cos\left(\sqrt{2 \beta} \dfrac{\phi}{M_{pl}} \right) \right\rbrace \right] \ .
\label{Potential1}
\ee
As for the scalar field, substituting (\ref{HubblePhi1}) in (\ref{dotphiH}) one finds
\begin{equation}
\phi = 2 \sqrt{\dfrac{2}{\beta}}\, M_{pl}\, \arctan(e^{\beta M t}) \ ,
\end{equation}
and plugging it back again into (\ref{HubblePhi1}) one can obtain the expression
\begin{equation}
H(t) = - M \tanh(\beta Mt) \ ,
\end{equation}
which implies that the scale factor behaves as
\begin{equation}
a \propto \cosh^{-1/\beta}(\beta M t) \ .
\end{equation}
It should be noted that the condition  $\beta>0$ guarantees that in order to produce inflation the potential  (\ref{Potential1}) will get to a minimum. Indeed, there is a critical value of the field for which $V(\phi_c)=0$, which corresponds to
\begin{equation}\label{critical}
\phi_c = \dfrac{M_{pl}}{\sqrt{2\beta}} \arccos\left(1 - \dfrac{6}{3+\beta}   \right) \ .
\end{equation}
In order to cut the potential before getting to negative values, we set the cut-off field $\phi_0$ as $\phi_0 < \phi_c$ in such a way that depending on how small this cut-off is as compared to the critical value (\ref{critical}) we find different inflationary behaviours. For instance, if $\phi_0 \ll \phi_c$ then the model is similar to a quadratic hilltop inflation with a cut-off, while if  $\phi_0 \lesssim \phi_c$ then the model resembles natural inflation with an additional negative cosmological constant, $ \Lambda = M^2 (3 + \alpha)$.

To compute the slow roll parameters we define a field position, $\phi_i$, as the one which is 55 e-folds back from $ \phi_c$. Next, we substitute Eq.\eqref{Potential1} into \eqref{sllowRollParam} to find
\begin{align}
&\epsilon \equiv \dfrac{1}{2} \left(\dfrac{V'}{V}\right)^2 = \dfrac{\beta (3+ \beta)^2 \sin^2(\sqrt{2\beta}/M_{pl})}{[-6-\alpha + \alpha \cos(\sqrt{2\beta}/M_{pl})]^2}  \\
&\eta \equiv \dfrac{V''}{V}= \dfrac{2\beta (3+ \beta) \cos(\sqrt{2\beta}/M_{pl})}{-3+\beta - (3 + \beta) \cos(\sqrt{2\beta}/M_{pl})} \ .
\end{align}
For the slow roll approximation to hold, we must constrain the largest values these parameters can take to be of order  $\mathcal{O} (10^{-2}) $ for $0,005 < \beta < 0,025$ and $0 < \phi < \phi_i$. Therefore, choosing different values for $\beta$ one gets different values for the slow roll parameters that can be substituted in the spectral parameters (\ref{spectralAndr}) and compared to  observational data.

\section{Inflation in multifield models}
\label{multifields}

Let us start by analyzing the main general features in inflation when considering more than one scalar field to produce the accelerating expansion. The general minimally coupled multifield scenario is described by action
\begin{equation}
S_E = \int d^4x \sqrt{-g} \left[\dfrac{M^2_{pl}}{2} R- \dfrac{1}{2} \mathcal{G}_{IJ} \, g^{\mu \nu} \, \partial_\mu \,\varphi^{I} \partial_\nu\, \varphi^{J}- V(\varphi^I) \right]\ ,
\end{equation}
where the $I,J$ indices run from $1$ to $n$, while the $n \times n$ metric $\mathcal{G}_{IJ}$ determines the kinetic terms in the fields space, which in the most general case may include non-canonical and crossed kinetic terms. In multifield scenarios, it is usual to define a new field usually known as the adiabatic field, that represents the path length along the classical trajectory, as \cite{Gordon:2000hv}
\begin{equation}
\dot{\sigma}^2 = \mathcal{G}_{IJ} \dot{\varphi}^I \dot{\varphi}^J \ .
\end{equation}
Then, the corresponding background FLRW equations are given by:
\begin{equation}
H^2= \dfrac{1}{3M^2_{pl}} \left[\dfrac{1}{2} \dot{\sigma}^2 + V \right] \ , \quad \dot{H} = - \dfrac{1}{2M^2_{pl}} \dot{\sigma}^2\ ,
\label{FLRWmultifieldEqs}
\end{equation}
while the equation for the adiabatic field $\sigma$ is given by:
\be
\ddot{\sigma} + 3H\dot{\sigma} + V_\sigma = 0\ .
\label{sigmaEq}
\ee
where $V_\sigma = \hat{\sigma}^I \dfrac{dV}{d\dot{\varphi}^I}$ with $\hat{\sigma}^I \equiv \dfrac{\dot{\varphi}^I_0 }{\dot{\sigma}}$, being $ \varphi^I_0 $ the background value. Now the background dynamics looks like a single field model with canonical kinetic terms, but one has to take care of the potential, since it depends on all the independent fields.
For simplicity, we are considering here two scalar fields $\{\phi, \chi\}$, although this analysis can be easily generalized to $n$ scalar fields, and furthermore we shall consider canonical kinetic terms (minimal couplings $\mathcal{G}_{IJ} = \delta_{IJ}$), so that the time derivative of the adiabatic field is:
\begin{equation}
\dot{\sigma}^2 = \dot{\phi}^2 + \dot{\chi}^2\ .
\label{adiabaticField}
\end{equation}
This expression can be rewritten in the following form:
\begin{equation}
\dot{\sigma} = \cos \theta \, \dot{\phi} + \sin \theta \, \dot{\chi}\ ,
\end{equation}
where $\cos \theta = \dfrac{\dot{\phi}}{\sqrt{\dot{\phi}^2 + \dot{\chi}^2}}$ and $\sin \theta = \dfrac{\dot{\chi}}{\sqrt{\dot{\phi}^2 + \dot{\chi}^2}}$. For the appropriate potential $V(\sigma)$, an accelerating expansion can be easily achieved similarly as in single field models. Nevertheless, the field trajectories play a fundamental role in the generation of non-adiabatic perturbations.

Let us consider perturbations of the scalar fields as $\delta\phi$ and $\delta\chi$, and consider the fluctuation on the entropy field, $s$, defined as
\begin{equation}
\delta s = \cos \theta \, \delta{\chi} -\sin \theta \, \delta{\phi} \ .
\end{equation}
Whenever both fields have equal trajectories in the background, the entropy field is null, $\delta s=0$. Here we consider scalar perturbations on the metric,
\begin{equation}
ds^2= - (1+2A)dt^2 + 2aB_i dx^i dt  + a^2 [(1-2\psi) \delta_{ij} + 2E_{ij}]dx^idx^j \ .
\end{equation}
Then, working in the spatially flat gauge, the gauge-invariant Mukhanov-Sasaki variable is given by:
\begin{align}
Q_\sigma \equiv \hat{\sigma}_I Q^I = \delta\sigma + \dfrac{\dot{\sigma}}{H} \psi\ ,
\end{align}
which accounts for the non-adiabatic fluctuation $\delta\sigma$. We may define the gauge-invariant curvature perturbation, $\mathcal{R}$, as:
\begin{equation}\label{curv.perturb}
\mathcal{R} \equiv \psi - \dfrac{H}{\rho + P} \delta q= \psi +\dfrac{H}{\dot{\sigma}} \hat{\sigma}_I \delta \phi ^I = \dfrac{H}{\dot{\sigma}} Q_\sigma \ ,
\end{equation}
where we have used the total momentum perturbation or the energy density flux of the perturbed fluid, $\delta q$, which is given by:
\begin{align}
\delta q = - \dot{\sigma} \hat{\sigma_J} \delta \varphi^J= - \dot{\sigma} Q\sigma \ ,
\end{align}	
while the total energy density, $\rho$, and pressure, $P$, are written as
\begin{align}
\rho = \dfrac{1}{2}\dot{\sigma}^2 + V\ , \quad P = \dfrac{1}{2}\dot{\sigma}^2 - V\ .
\end{align}
Then, by using the perturbed FLRW equations and the scalar fluctuation field equation, the evolution equation for the field perturbations can be expressed as \cite{Gordon:2000hv}
\begin{equation}
\ddot{Q_\sigma} + 3H \dot{Q_\sigma} + \left[\dfrac{k^2}{a^2} + V_{\sigma\sigma} - \dot{\theta}^2 - \dfrac{1}{M^2_{pl} a^3} \dfrac{d}{dt} \left(\dfrac{a^3 \dot{\sigma}^2}{H}\right) \right] Q_\sigma = 2 \dfrac{d}{dt} (\dot{\theta}\delta s) - 2 \left( \dfrac{V_\sigma}{\dot{\sigma}} + \dfrac{\dot{H}}{H}\right) \dot{\theta} \delta s\ ,
\label{adiab.perturb}
\end{equation}	
and
\begin{equation}
\ddot{\delta s} + 3H\dot{\delta s} + \left(\dfrac{k^2}{a^2} + V_{ss} + 3\dot{\theta}^2  \right) \delta s = \dfrac{4 M^2_{pl}k^2 \dot{\theta}}{\dot{\sigma}a^2} \Psi\ ,
\label{entropy.perturb}
\end{equation}
where $ \Psi $ is the gauge-invariant Bardeen potential $\Psi \equiv \psi + a^2 H \left( \dot{E}-\dfrac{B}{a}\right)$, $ \dot{\theta} =- \dfrac{V_s}{\dot{\sigma}}$ with $V_s$ being the potential gradient perpendicular to the trajectory in the field space:
\begin{equation}\label{key}
V _s = \cos\theta \; V_\chi - \sin\theta \; V_\phi \ .
\end{equation}
As shown in \eqref{adiab.perturb}, the entropy perturbation $\delta s$ works as a source term for adiabatic perturbations. Taking the time derivative of the curvature perturbation (\ref{curv.perturb}) one finds
\begin{equation}\label{deriv.curv.perturb}
\dot{\mathcal{R}} = \dfrac{H}{\dot{H}} \dfrac{k^2}{a^2} \Psi + \dfrac{2H}{\dot{\sigma}} \dot{\theta} \delta s \ .
\end{equation}
We can see that $\mathcal{R} $ is not conserved even in the large-scale limit whenever $ \delta s \neq 0$ and a non-straight trajectory in the field space ($\dot{\theta} \neq 0$) occurs \cite{Bassett:2005xm}.
The dimensionless power spectrum is given by:
\begin{equation}
\mathcal{P}_\mathcal{R} (k) = \dfrac{k^2}{2\pi^2}|\mathcal{R}|^2\ ,
\end{equation}
whereas the spectral index is defined as follows:
\begin{equation}
n_s= 1 + \dfrac{\partial \ln \mathcal{P}_\mathcal{R}}{\partial\ln k}\ .
\end{equation}
Hence, the non-adiabatic perturbations might induce deviations on the spectral index in such a way that when assuming the presence of more than a scalar field in inflation the corresponding trajectories in the field space play a fundamental role, since such corrections may provide wrong predictions on the spectral index (and on the tensor to scalar ratio as well) when comparing to the Planck data. In the next section, we analyze a two scalar field model when both fields constant roll.

\section{Contant roll Inflation with two scalar fields}
\label{multifieldsConstant}

Here we consider a two scalar field inflation model, whose field equations (\ref{FLRWmultifieldEqs}) are written as
\begin{eqnarray}
3M_{Pl}^2&=&\frac{1}{2} (\dot{\phi}^2 + \dot{\chi}^2) + V(\phi,\chi)\ , \nn
-2M_{Pl}^2 \dot{H}&=&\dot{\phi}^2 + \dot{\chi}^2 \ , \label{eq:mfeq2}
\end{eqnarray}
together with the respective equations for each field, that is,
\be
\ddot{\phi}+3H\dot{\phi}+V_{\phi}=0\  \quad \text{and} \quad \ddot{\chi}+3H\dot{\chi} +V_{\chi}=0 \ .
\label{scalarFieldsEqs}
\ee
As in the case of single field model, we impose constant roll conditions on the two fields, which are written as:
\begin{equation}
\frac{\ddot{\phi}}{\dot{\phi}}=\beta_{\phi} H\ , \quad  \frac{\ddot{\chi}}{\dot{\chi}}=\beta_{\chi} H\ ,
\label{CRconditionMulti}
\end{equation}
where $\beta_{\phi}$ and $\beta_{\chi}$ are constants. In order to carry out our analysis, we must distinguish the cases in which these two constants are equal or different.

\subsection{$\beta_{\phi}= \beta_{\chi}$}
\label{multifieldsConstant1}
\begin{figure}[t]
   \centerline{ \includegraphics[width=10cm,height=6cm]{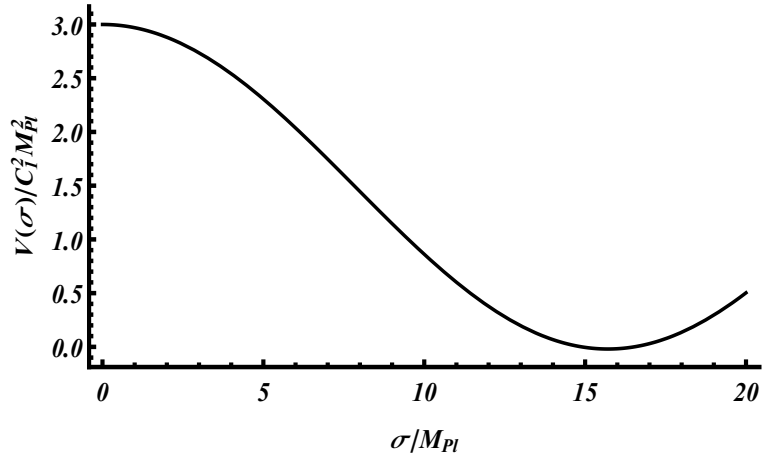}}
\caption{The potential $V(\sigma)$ in Eq.(\ref{Vsigma2}) in units of $C_1^2M_{Pl}^2$ for $\beta=0.02$. The corresponding critical value $V(\sigma_c)=0$ is given by $\sigma_c/M_{Pl}=14.9$.}
  \label{Fig:Potential1}
\end{figure}

Let us start by considering $\beta_{\phi}= \beta_{\chi}=\beta$. In such a case, by the constant roll conditions (\ref{CRconditionMulti}) one has:
\be
\beta H=\frac{\ddot{\phi}}{\dot{\phi}}=\frac{\ddot{\chi}}{\dot{\chi}}\ .
\label{CRconditionMultiCase1}
\ee
The second part of this equation can be integrated in such a way that both fields become related as
\be
\dot{\phi}=k_1\dot{\chi}\ \rightarrow \phi=k_1 \chi + k_0\ ,
\label{phiChiRel1}
\ee
with $k_0,k_1$ integration constants. Then, by the scalar field equation for $\phi$ in (\ref{scalarFieldsEqs}), the following extra relation is obtained:
\be
\ddot{\phi}+3H\dot{\phi}+\frac{\partial V}{\partial\phi}=k_1\ddot{\chi}+3k_1H\dot{\chi}+\frac{1}{k_1}\frac{\partial V}{\partial\chi}=0\ ,
\label{phiChiRel2}
\ee
By comparing the last part of this expression with the field equation for $\chi$ in (\ref{scalarFieldsEqs}), a constraint on $k_1$ is obtained:
\be
k_1=\pm 1\ .
\label{condk1}
\ee
Our aim now is to reconstruct the most general potential $V(\phi, \chi)$ that holds the constant roll condition (\ref{CRconditionMulti}). From the second FLRW equation (\ref{eq:mfeq2}) and using (\ref{phiChiRel1}), one can compute the left-hand side as $-2M_{Pl}^2 \dot{H}=-4M_{Pl}^2\frac{\partial H}{\partial \phi}$, so that by comparing both sides of Eq.(\ref{eq:mfeq2}) we find
\begin{equation}
\dot{\phi}=-\frac{4M_{Pl}}{1+k_1^{-2}}\frac{\partial H}{\partial \phi} \ ,
\end{equation}
By applying the time derivative here, the following relation is obtained:
\begin{equation}
\frac{\ddot{\phi}}{\dot{\phi}}=-\frac{8M_{Pl}}{1+k_1^{-2}}\frac{\partial^2 H}{\partial \phi^2}\ .
\end{equation}
Therefore, by the constant roll condition (\ref{CRconditionMulti}), an equation for $H$ in terms of the scalar field $\phi$ is obtained:
\begin{equation}
\frac{\partial^2 H}{\partial \phi^2}+\frac{\beta}{8M_{Pl}^2}\left(1+k_1^{-2}\right)H=0\ ,
\label{HphiEq1}
\end{equation}
which is the natural generalization of (\ref{eqHphi}). By solving (\ref{HphiEq1}) one obtains:
\bea
H(\phi)&=C_1 \exp\left(\frac{1}{2}\sqrt{-\frac{\beta(1+k_1^2)}{2k_1^2}}\frac{\phi}{M_{Pl}}\right)+C_2 \exp\left(-\frac{1}{2}\sqrt{-\frac{\beta(1+k_1^2)}{2k_1^2}}\frac{\phi}{M_{Pl}}\right) \ .
\label{Hphi1}
\eea
The potential $V(\phi,\chi)=V(\phi, \phi/k_1-k_0)=V(\phi)$ can be reconstructed by using this result in the first FLRW equation (\ref{eq:mfeq2}), leading to
\be
V(\phi)=M_{Pl}^2 \left[C_1^2(3+\beta) \exp\left(\sqrt{-\frac{\beta(1+k_1^2)}{2k_1^2}}\frac{\phi}{M_{Pl}}\right)+C_2^2(3+\beta) \exp\left(-\sqrt{-\frac{\beta(1+k_1^2)}{2k_1^2}}\frac{\phi}{M_{Pl}}\right)-2C_1C_2(\beta-3)\right]\ .
\label{PotentialCase1}
\ee
Now, from the relation between the scalar fields (\ref{phiChiRel1}) and the integrability condition on the potential $\frac{\partial^2V}{\partial\phi\partial\chi}=\frac{\partial^2V}{\partial\chi\partial\phi}$, one gets $k_0=0$ and the full potential in terms of both scalar fields becomes
\be
V(\phi, \chi)=M_{Pl}^2 \left[C_1^2(3+\beta) \exp\left(\frac{\sqrt{-\beta(\phi^2+\chi^2)}}{M_{Pl}}\right)+C_2^2(3+\beta) \exp\left(-\frac{\sqrt{-\beta(\phi^2+\chi^2)}}{M_{Pl}}\right)-2C_1C_2(\beta-3)\right]\ .
\label{PotentialCase1}
\ee
Depending on the value of $\beta$, the nature of the potential will be different and, consequently, so will be the way inflation occurs. Nevertheless, it is more convenient to work with the adiabatic field $\sigma$ and the entropic field $s$ as defined in the above section. For this purpose, let us start by analyzing the behaviour of the adiabatic field (\ref{adiabaticField}) under the constant roll conditions (\ref{CRconditionMultiCase1}):
\be
\ddot{\sigma}=\frac{\dot{\phi}\ddot{\phi}+\dot{\chi}\ddot{\chi}}{\sqrt{\dot{\phi}^2+\dot{\chi}^2}}=\beta H\sqrt{\dot{\phi}^2+\dot{\chi}^2}=\beta H\dot{\sigma} \quad \rightarrow\quad \frac{\ddot{\sigma}}{\dot{\sigma}}=\beta H\ .
\label{ConstantRollSigma}
\ee
Therefore the adiabatic field also constant rolls. We can use the FLRW equations (\ref{FLRWmultifieldEqs}) and the scalar field equation (\ref{sigmaEq}) to write the analog equation (\ref{HphiEq1}) for $H$ as a function of the adiabatic field as
\be
 \frac{\partial^2 H}{\partial \sigma^2}+\frac{\beta}{2M_{Pl}^2}H=0\ ,
 \label{sigmaHeq}
 \ee
 whose solution is
 \be
 H(\sigma)=C_1\exp\left(\sqrt{\frac{-\beta}{2}}\frac{\sigma}{M_{Pl}}\right)+C_2\exp\left(-\sqrt{\frac{-\beta}{2}}\frac{\sigma}{M_{Pl}}\right)\ ,
 \label{Hsigma}
 \ee
while the constant roll potential leads to
\be
V(\sigma)=M_{Pl}^2\left[C_1^2(3+\beta)\exp\left(\sqrt{-2\beta}\frac{\sigma}{M_{Pl}}\right)+C_2^2(3+\beta)\exp\left(-\sqrt{-2\beta}\frac{\sigma}{M_{Pl}}\right)-2C_1C_2(\beta-3)\right]\ .
\label{Vsigma}
\ee
Hence, we get the potential that describes the constant roll adiabatic field $\sigma$. As pointed out in Ref.~\cite{Motohashi:2014ppa}, the only viable potential for inflation corresponds to values $\beta>0$, leading to:
\be
V(\sigma)=\frac{1}{2}M_{Pl}^2C_1^2\left[3-\beta+(3+\beta)\cos\left(\sqrt{2\beta}\frac{\sigma}{M_{Pl}}\right)\right]\ ,
\label{Vsigma2}
\ee
which is depicted in Fig.\ref{Fig:Potential1} for $\beta=0.02$.
The corresponding solutions for the Hubble parameter and the scalar field can be obtained from the FLRW equations (\ref{FLRWmultifieldEqs}) as
\be
H(t)=-C_1\tanh\left(C_1\beta t\right) \ , \quad \sigma(t)=2\sqrt{\frac{2}{\beta}}M_{Pl}\arctan\left(\e^{C_1\beta t}\right)\ .
\label{sol1}
\ee
Nevertheless, it is more convenient to write the solutions in terms of the number of e-foldings, which can be obtained from the second FLRW equation (\ref{FLRWmultifieldEqs}) expressed as follows
\be
H\frac{\partial\sigma}{\partial N}2M_{Pl}^2+\frac{\partial H}{\partial \sigma}=0\ .
\label{FLRWNfolds}
\ee
Using the scalar potential (\ref{Vsigma2}) and the first FLRW equation in (\ref{FLRWmultifieldEqs}), the following solutions for $H$ and $\sigma$ as functions of the number of e-foldings are obtained
\be
H(N)=C_1\sqrt{1-\exp\left(2\beta N+m\sqrt{2\beta}\right)}\ , \quad \sigma(N)=\sqrt{\frac{2}{\beta}}M_{Pl}\arcsin\left(\beta N+m\sqrt{\frac{\beta}{2}}\right)\ .
\label{HsigmaNfolds}
\ee
Here $m$ is a constant that can be fixed with the number of e-foldings that inflation lasts, as shown below. As in the case of a single field model, the potential (\ref{Vsigma2}) should be cut off prior to a critical value where $V(\sigma_c)=0$, that is
\be
\sigma_c=\frac{M_{Pl}}{\sqrt{2\beta}}\arccos\left(\frac{\beta-3}{\beta+3}\right)\ .
\label{sigmac}
\ee
As we are interested in evaluating the perturbations at a field point $\sigma_i<\sigma_c$, this can be fixed a number of e-foldings $N_c$ before $\sigma_c$, which in general amounts to $50-65$. Then, using (\ref{HsigmaNfolds}) and (\ref{sigmac}), the constant $m$ is expressed as:
\be
m=\frac{1}{\sqrt{2\beta}}\left[-2\beta N_c+\log\left(\frac{3}{3+\beta}\right)\right] \ .
\label{constantm}
\ee
As shown in Fig.~\ref{Fig:Potential1}, inflation occurs prior to the critical value $\phi_c$. Before analyzing the perturbations, we should study the non-adiabatic perturbations in order to compute the possible contributions to the spectral index. Nevertheless, as the first derivative for both fields are proportional $\dot{\phi}=k_1\dot{\chi}$, the fluctuation $\delta s$ becomes null:
\be
\delta s=\cos \theta \delta\chi-\sin\theta \delta\phi=(\cos\theta-\sin\theta)\delta\chi=\left(\frac{k_1}{\sqrt{1+k_1^2}}-\frac{k_1}{\sqrt{1+k_1^2}}\right)\delta\chi=0\ .
\label{deltasCase1}
\ee
Hence, the spectral index $n_s$ and the tensor to scalar ratio $r$ can be computed through (\ref{spectralAndr}), where the slow roll parameters are now given by:
\be
\epsilon=\frac{M_{Pl}^2}{2}\left(\frac{V_{\sigma}}{V}\right)^2\ ,\quad \eta=M_{Pl}^2\frac{V_{\sigma\sigma}}{V}\ .
\label{sllowRollParamSigma}
\ee
 \begin{figure}[t]
  \includegraphics[scale=0.6]{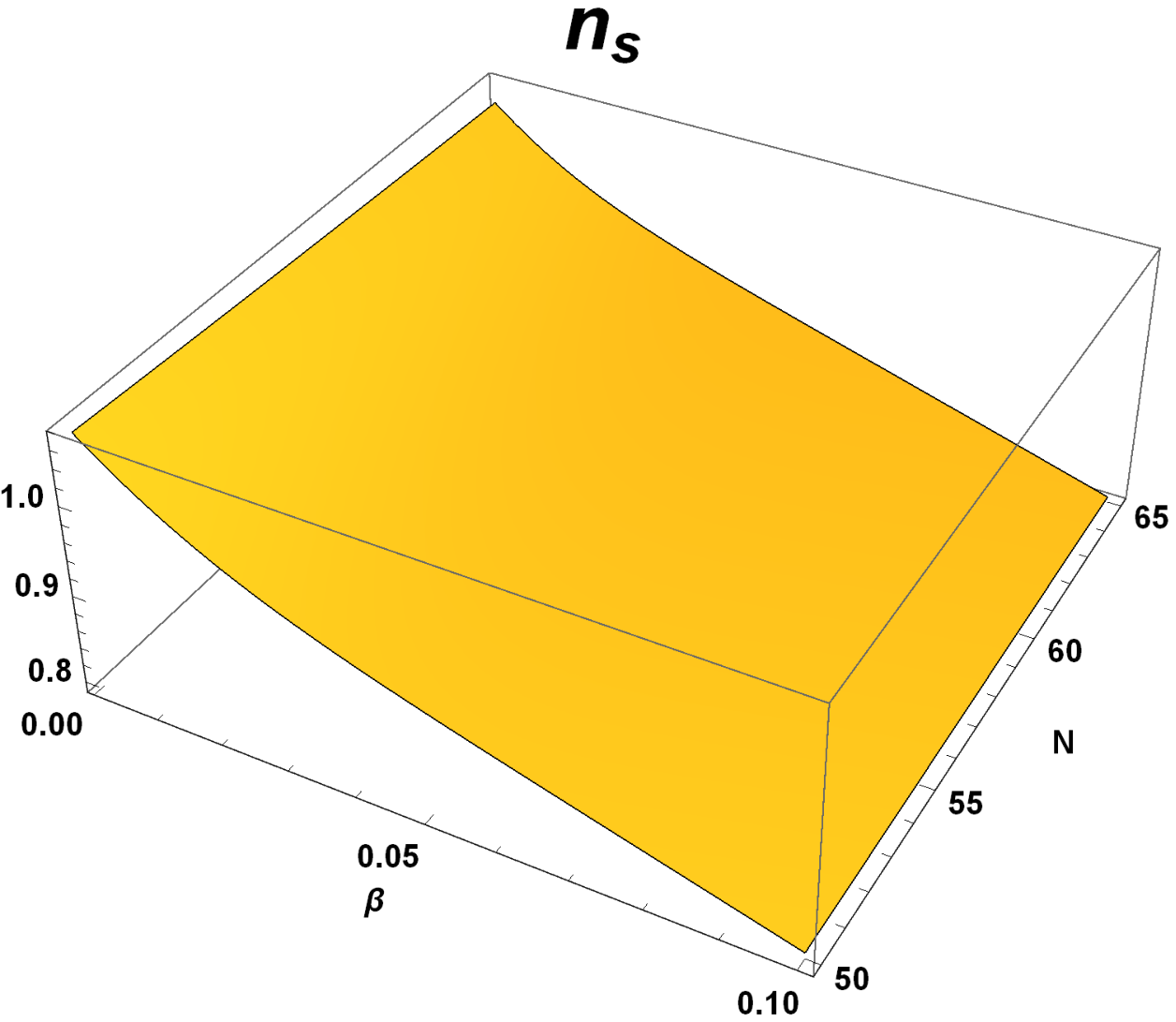}
 \includegraphics[scale=0.6]{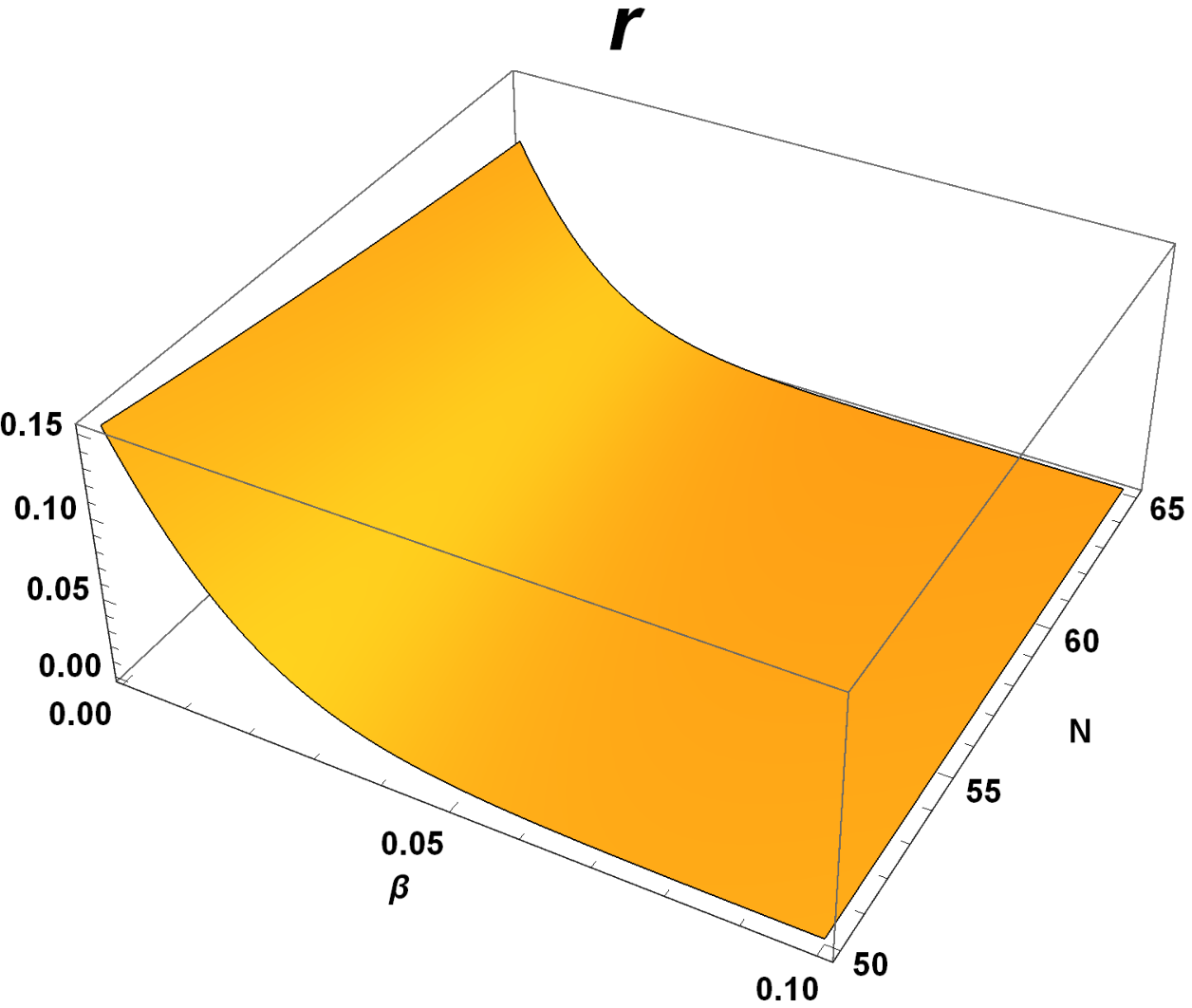}
\caption{Spectral index $n_s$ and tensor to scalar ratio $r$ for the case $\beta_{\phi}=\beta_{\chi}=\beta$ as a function of the number of e-foldings $N=50-65$ and of $\beta=0-0.10$.}
  \label{spectral_indexFig}
\end{figure}
 \begin{figure}[t]
   \centerline{ \includegraphics[scale=0.6]{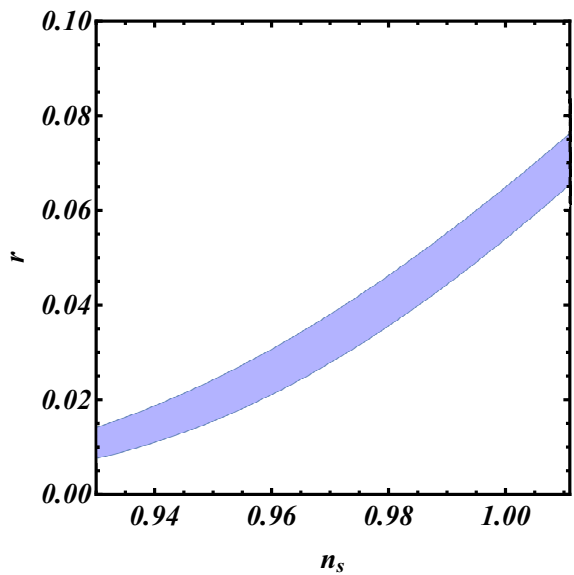}}
\caption{Parametric plot for the spectral index $n_s$ versus the tensor to scalar ratio $r$.}
  \label{nsr}
\end{figure}

In Fig.~\ref{spectral_indexFig}, the spectral index $n_s$ and the tensor to scalar ratio $r$ are depicted as functions of the parameter $\beta$ and the number of e-foldings that $\sigma_i$ is located back from the critical value $\sigma_c$. On the other hand, Fig.~\ref{nsr} shows the parametric plot for both quantities, where the allowed values region is depicted. One should bear in mind that the last constraints on both parameters provided by Planck \cite{Akrami:2018odb,Ade:2015lrj} are:
\be
n_s=0.9659\pm 0.0041\ , \quad r<0.11\ .
\label{constraintsPlanck}
\ee
Hence, as shown in Figs.~\ref{spectral_indexFig} and \ref{nsr}, the constant roll model with two fields that rolls down at the same rate (\ref{CRconditionMultiCase1}), but may not in the same way, predicts an spectral index and tensor to scalar ratio that can fit the Planck constraints, although the value of $r$ may excess the upper constraint (\ref{constraintsPlanck}) in the confidence region for $n_s$.

\subsection{$\beta_{\phi}\neq \beta_{\chi}$}
\label{multifieldsConstant2}

\begin{figure}[t]
  \includegraphics[scale=0.6]{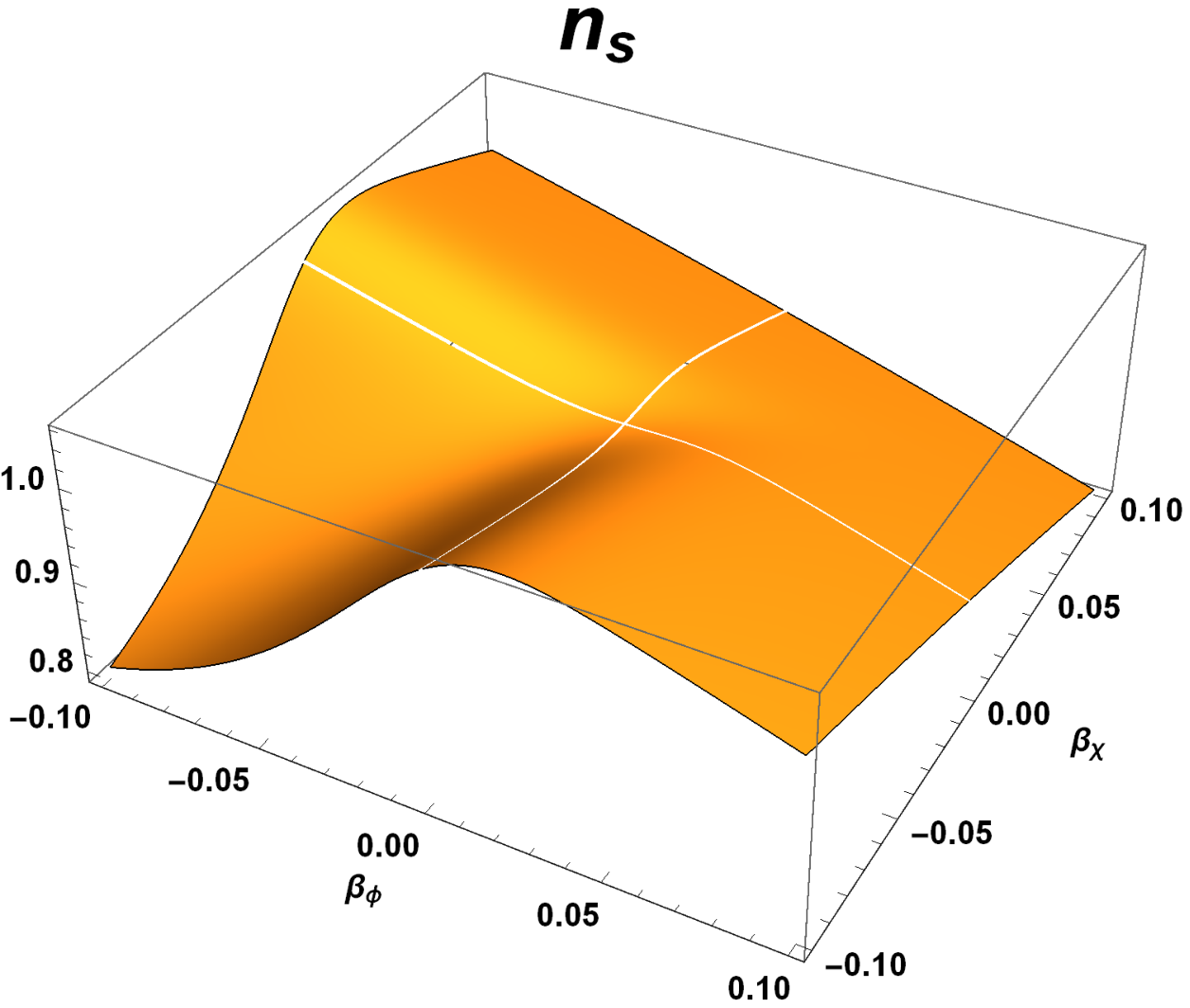}
 \includegraphics[scale=0.6]{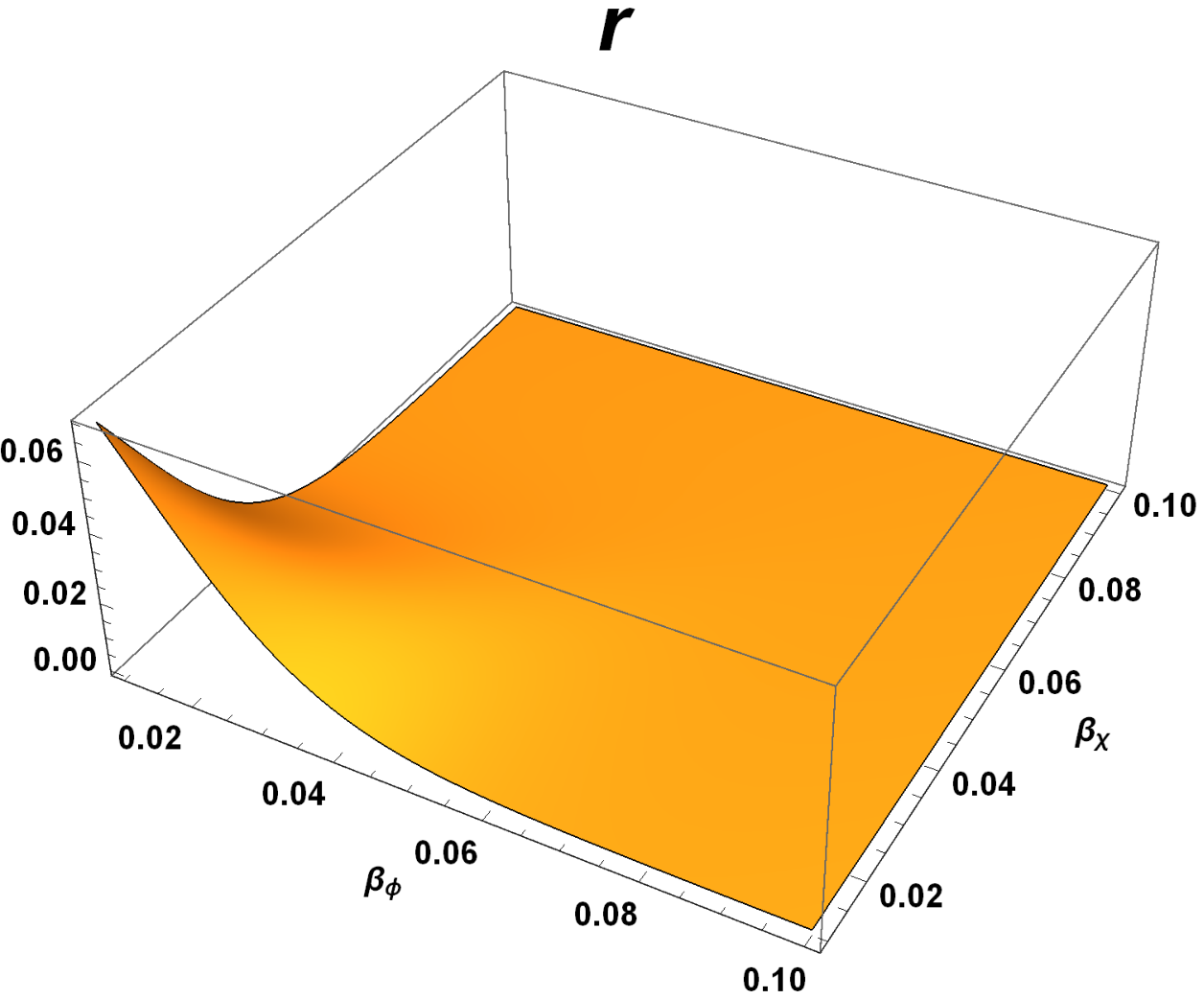}
   \includegraphics[scale=0.6]{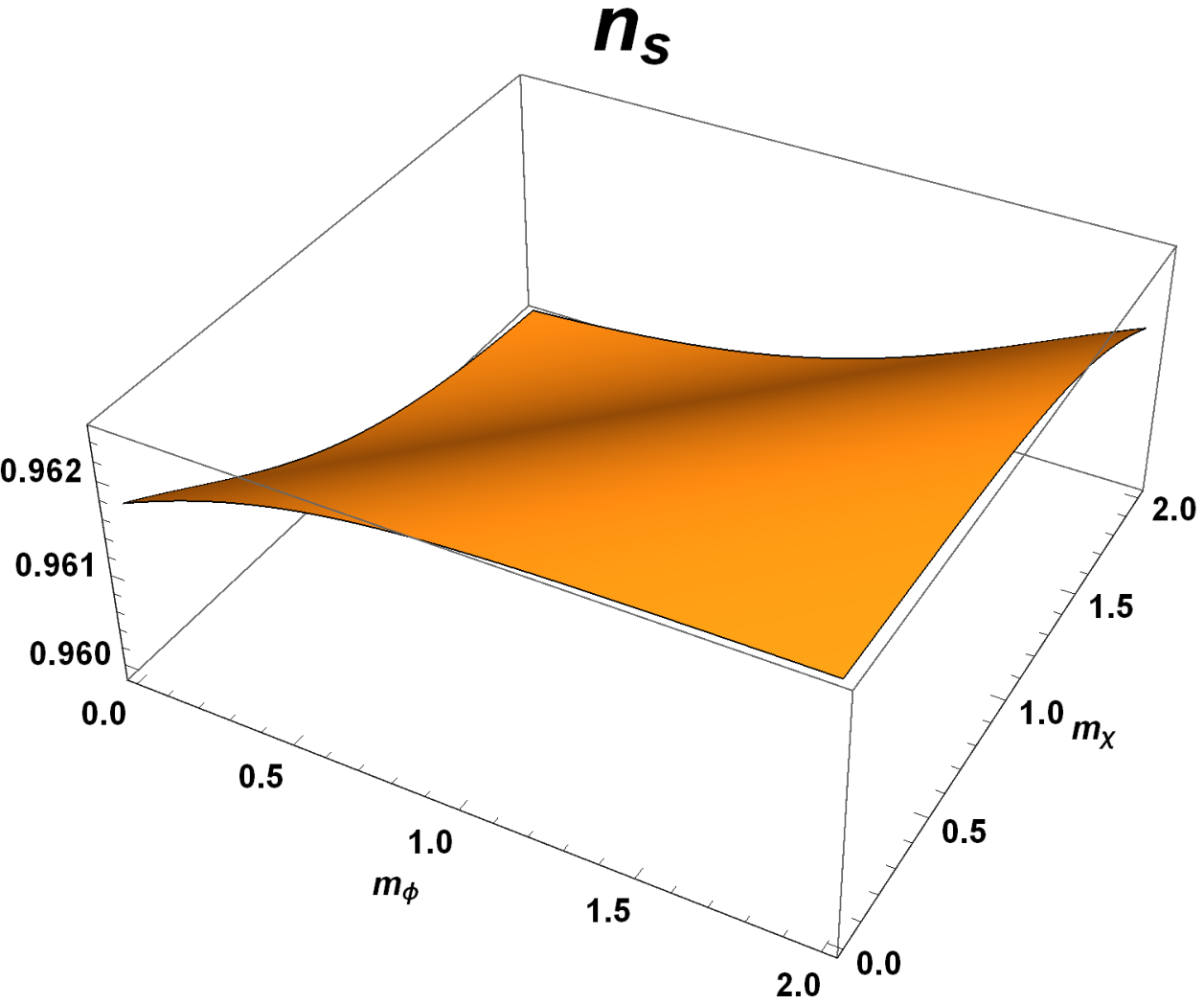}
  \includegraphics[scale=0.6]{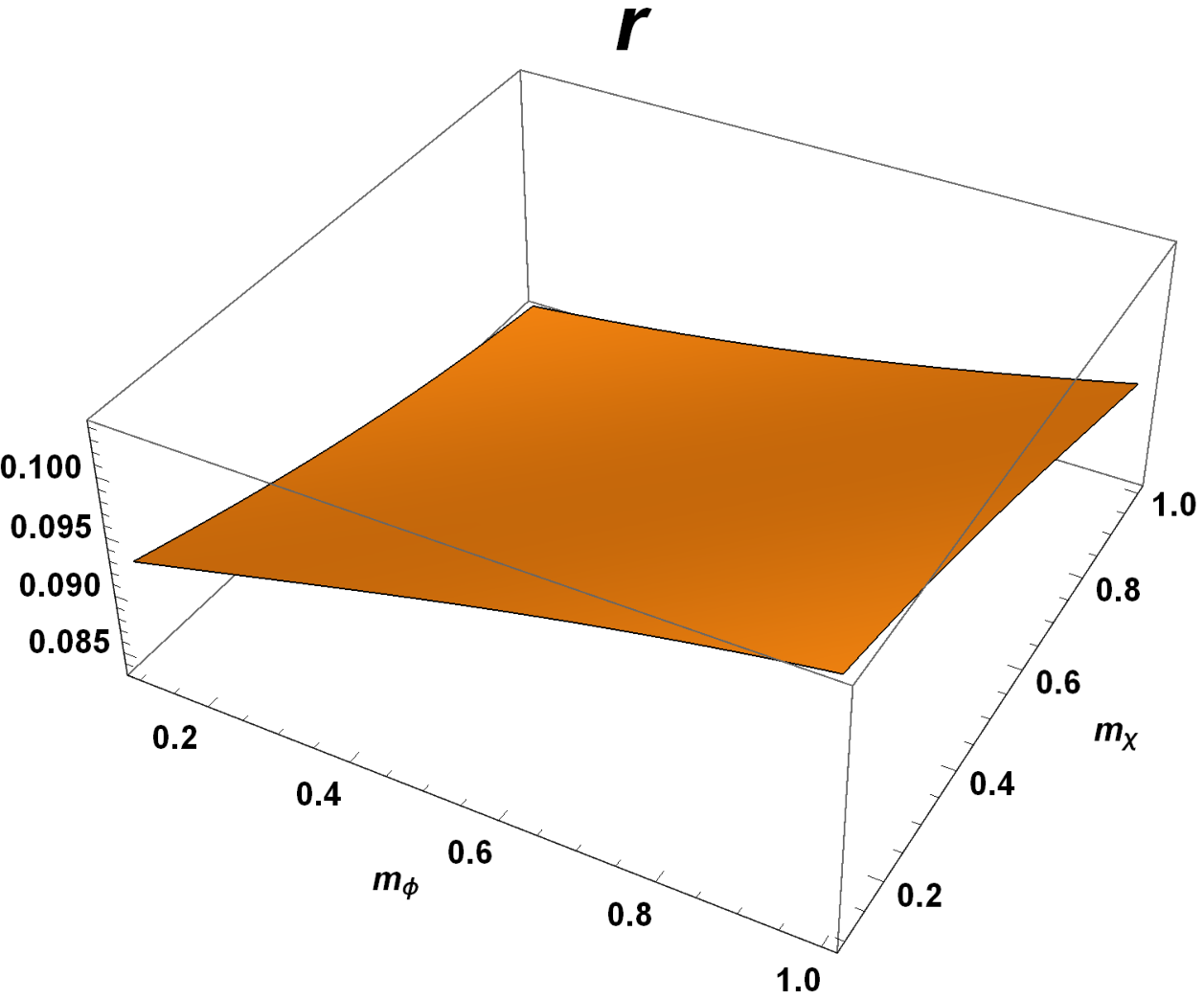}
\caption{Spectral index $n_s$ and tensor to scalar ratio $r$ for the case $\beta_{\phi}\neq\beta_{\chi}=\beta$. Top panels show the variation of both magnitudes with respect to the constant roll parameters, while in bottom panels the same magnitudes are depicted as functions of the initial velocities for the scalar fields. Both cases are considered for a 55 e-foldings inflation.}
  \label{spectral_index_r_Fig4}
\end{figure}

Let us now consider the case where both fields constant roll at different rates, i.e., $\beta_{\phi}\neq\beta_{\chi}$. As in the case above, by the constant roll conditions (\ref{CRconditionMulti}), we can relate both fields as follows:
\be
\frac{1}{\beta_{\phi}}\frac{\ddot{\phi}}{\dot{\phi}}=\frac{1}{\beta_{\chi}}\frac{\ddot{\chi}}{\dot{\chi}}\ \rightarrow \dot{\phi}\propto (\dot{\chi})^{\beta_{\phi}/\beta{\chi}}\ .
\label{constantRolldiff}
\ee
It is clear that for very different constant roll parameters, for instance $\beta_{\chi}>>\beta_{\phi}$, the problem reduces to one single field, similarly to the case analyzed above. For the general case, it is not possible to reconstruct the corresponding potential $V(\phi, \chi)$ analytically, but it can be computed in terms of the number of e-foldings. Indeed, by integrating independently the constant roll conditions (\ref{CRconditionMulti}), one obtains:
\bea
\dot{\phi}=M_{Pl}\dot{\phi}_{0}\left(\frac{a}{a_0}\right)^{\beta_{\phi}}=M_{Pl}\e^{\beta_{\phi}N+m_{\phi}}\ , \nn 
 \dot{\chi}=M_{Pl}\dot{\chi}_{0}\left(\frac{a}{a_0}\right)^{\beta_{\chi}}=M_{Pl}\e^{\beta_{\chi}N+m_{\chi}}\ ,
\label{phichidotN}
\eea
where $m_{\phi}$ and $m_{\chi}$ are integration constants that determine the initial velocities of the scalar fields $\dot{\phi}_0$ and $\dot{\chi}_0$. The second FLRW equation (\ref{eq:mfeq2}) can be expressed in terms of the number of e-foldings as follows:
\be
-2H H'=\e^{2\beta_{\phi}N+2m_{\phi}}+\e^{2\beta_{\chi}N+2m_{\chi}}\ ,
\label{HefoldsEq2}
\ee
whose solution is given by:
\be
H(N)=\sqrt{2C_1-\frac{\e^{2\beta_{\phi}N+2m_{\phi}}}{2\beta_{\phi}}-\frac{\e^{2\beta_{\chi}N+2m_{\chi}}}{2\beta_{\chi}}}\ ,
\label{HefoldsSol2}
\ee
with $C_1$ an integration constant. A glance to this expression allows to realize that the Hubble parameter seems to behave similarly as in the case of equal constant roll, as given by Eq.(\ref{HsigmaNfolds}). The corresponding potential $V(\phi, \chi)=V(N)$ can be obtained from the first FLRW equation (\ref{eq:mfeq2}) as
\be
V(N)=\frac{M_{Pl}^2}{2}\left(\frac{3+\beta_{\phi}^2}{\beta_{\phi}}\e^{2\beta_{\phi}N+2m_{\phi}}+\frac{3+\beta_{\chi}^2}{\beta_{\chi}}\e^{2\beta_{\chi}N+2m_{\chi}}-12C_1\right)\ .
\label{PotentialNCase2}
\ee
The corresponding $\phi(N)$ and $\chi(N)$ fields could then be obtained by integrating (\ref{phichidotN}), but this path would not provide an exact expression, and therefore the potential cannot be obtained for this general case. Nevertheless, similarly as in the previous case, we can analyze the behaviour of the adiabatic field $\sigma$ and the entropy perturbations $\delta s$ to extract information on how the fluctuations are generated when $\beta_{\phi}\neq \beta_{\chi}$. Let us start by analyzing the behaviour of $\sigma$ as defined in (\ref{adiabaticField}), which in the present case reads explicitly
\bea
\dot{\sigma}&=&\sqrt{\dot{\phi}^2+\dot{\chi}^2}=\dot{\chi}\sqrt{1+k_1^2\dot{\chi}^{2\left(\frac{\beta_{\phi}}{\beta_{\chi}}-1\right)}}\ , \nn
\ddot{\sigma}&=&\frac{1}{\sqrt{\dot{\phi}^2+\dot{\chi}^2}}\left(\beta_{\phi}H\dot{\phi}^2+\beta_{\chi}H\dot{\chi}^2\right)=\frac{\beta_{\chi}H\dot{\chi}}{\sqrt{1+k_1^2\dot{\chi}^{2\left(\frac{\beta_{\phi}}{\beta_{\chi}}-1\right)}}}\left(1+k_1^2\frac{\beta_{\phi}}{\beta_{\chi}}\dot{\chi}^{2\left(\frac{\beta_{\phi}}{\beta_{\chi}}-1\right)}\right)\ ,
\label{sigmaddot2}
\eea
where $k_1$ is a proportional constant between the first derivatives of the scalar fields. As pointed above, from here we can see that unless both constant parameters are similar, or one very large in comparison to the other, the adiabatic field $\sigma$ will not constant roll. In addition, the entropy fluctuations field for the general case yields here:
\be
\delta s=\frac{\dot{\phi}\delta\chi-\dot{\chi}\delta\phi}{\sqrt{\dot{\phi}^2+\dot{\chi}^2}}\simeq\frac{k_1}{\sqrt{1+k_1^2\dot{\chi}^{2\left(\frac{\beta_{\phi}}{\beta_{\chi}}-1\right)}}}\left(\dot{\chi}^{2\left(\frac{\beta_{\phi}}{\beta_{\chi}}-1\right)}\delta\chi-\int dt\ \frac{\beta_{\phi}}{\beta_{\chi}}\dot{\chi}^{\frac{\beta_{\phi}}{\beta_{\chi}}-1}\dot{\delta\chi}\right)\ .
\label{entropyfluc2}
\ee
Then, non-adiabatic perturbations will arise in the general case, as occurs in slow roll inflation with multifields. Nevertheless, we can consider such perturbations small outside the horizon \cite{Weinberg:2004kf}. Hence, we can compute  the spectral index and the tensor to scalar ratio through the slow roll parameters as given in (\ref{sllowRollParamSigma}) by using the potential (\ref{PotentialNCase2}). As in the above case, we should impose also a cut-off in the potential, which can be taken as $V(\phi_c, \chi_c)=V(N_c)=0$, where $N_c$ is the number of e-foldings complete along the inflationary period, which allowed us to fix the integration constant $C_1$. In Fig.~\ref{spectral_index_r_Fig4}, the spectral index $n_s$ and the tensor to scalar ratio $r$ are depicted as functions of the constant roll parameters $\{\beta_{\phi}, \beta_{\chi}\}$ (top panels) and the initial conditions for the velocities of the scalar fields $\{m_{\phi}, m_{\chi}\}$ (bottom panels). We have considered for both cases a inflationary expansion of 55 e-foldings, but we point out that other durations provide similar results. Note also that in this case we have allowed the constant roll parameters $\{\beta_{\phi}, \beta_{\chi}\}$ to take negative values. Hence, the corresponding predictions for $n_s$ and $r$ can fit well the values provided by Planck (\ref{constraintsPlanck}).

\section{Conclusion}
\label{conclusiones}

In the present work we have analyzed a two field inflationary model when both fields constant roll, extending previous analysis of this class of inflationary scenarios to the multifield case. As is done in multifield inflation, we have used the approach of redefining the scalar fields by using the so-called adiabatic field and the entropy field, which gather the adiabatic perturbations and the entropy ones, respectively. Then, we have separately analyzed two cases, one when both scalar fields have the same constant roll parameter and the other one when they are different. In both cases, we obtain the spectral index and the tensor to scalar ratio and compared them to the latest data from Planck.

For the first case, where the constant roll ratio is the same for both fields, we show that the adiabatic field also constant rolls while non-adiabatic fluctuations are null. The corresponding potential for the adiabatic field is obtained, as well as its expression in terms of both scalar fields, which showed the same behaviour as in the single field scenario studied in \cite{Motohashi:2014ppa,Motohashi:2017aob}. The exact solution for the Hubble parameter is also obtained in terms of the cosmic time and the number of e-foldings. Then, the spectral index and the tensor to scalar ratio are obtained, which remain as functions of the number of e-foldings that inflation lasts and the constant roll parameter. As shown in Figs.~\ref{spectral_indexFig} and \ref{nsr}, the constraints provided by the Planck mission can be well satisfied. 

The second case leads to a more complex approach, as the differences between the constant roll parameters of both fields do not lead to an adiabatic field that also constant rolls, while the entropy fluctuations are in general not null. Nevertheless, for large differences of both constant roll parameters, the problem can be reduced to the one of a single field. In the general case, we have indeed obtained the exact solution for the Hubble parameter as a function of the number of e-foldings and the corresponding potential is also obtained in terms of this independent variable. That allowed us to compute the spectral index and the tensor to scalar ratio as functions of the constant roll parameters, the initial  velocities of each field and the number of e-foldings. By considering negligible entropy perturbations the results were also compared to the Planck data and depicted in Fig.~\ref{spectral_index_r_Fig4}, which can be satisfied for some values of the free parameters, as in the case above.

Hence, the paper has presented an extension of multifield scenarios when considering two constant roll scalar fields. Results fit well the observational constraints and keep the entropy fluctuations small for some of the cases, leading to a healthy generalization of constant roll inflation to multifield models.

\section*{Acknowledgements}

MG is funded by the predoctoral contract 2018-T1/TIC-10431.
DRG is funded by the \emph{Atracci\'on de Talento Investigador} programme of the Comunidad de Madrid (Spain) No. 2018-T1/TIC-10431, and acknowledges further support from the Ministerio de Ciencia, Innovaci\'on y Universidades (Spain) project No. PID2019-108485GB-I00/AEI/10.13039/501100011033, the Funda\c{c}\~ao para a Ci\^encia e a Tecnologia (FCT, Portugal) research projects Nos. PTDC/FIS-OUT/29048/2017 and PTDC/FIS-PAR/31938/2017, the projects FIS2017-84440-C2-1-P (MINECO/FEDER, EU) and H2020-MSCA-RISE-2017 Grant FunFiCO-777740 and the Edital 006/2018 PRONEX (FAPESQ-PB/CNPQ, Brazil) Grant No. 0015/2019.
DS-CG is funded by the University of Valladolid (Spain) and by the program MOVILIDAD INVESTIGADORES UVa-BANCO SANTANDER 2020, and thanks the Department of Theoretical Physics and IPARCOS at Complutense University of Madrid for their hospitality while doing this work. This article is based upon work from COST Actions CA15117 and CA18108, supported by COST (European Cooperation in Science and Technology).

\end{document}